1

# Performance Enhancement of MVDC Aircraft Cables Using Micro-Multilayer Insulation Under Low-Pressure Conditions

Saikat Chowdhury, *Member, IEEE*, and Mona Ghassemi, *Senior Member, IEEE*

*Abstract*— The development of medium-voltage direct current (MVDC) cable systems for wide-body all-electric aircraft (AEA) requires insulation technologies capable of operating reliably under reduced-pressure environments. Conventional underground cable insulation, designed for atmospheric conditions, exhibits degraded partial discharge (PD) and dielectric performance at low pressure, limiting its applicability to aerospace systems. This work presents a controlled experimental comparison between a conventional single-layer extruded insulation system and a micro-multilayer multifunctional electrical insulation (MMEI) architecture, in which all cable components are kept identical except for the insulation. The MMEI system is implemented with only 10% of the baseline insulation thickness to evaluate the effectiveness of insulation architecture in enhancing performance. PD characteristics and dielectric strength are experimentally evaluated under DC voltage at atmospheric pressure and 18.8 kPa. Results show that the MMEI-based cable exhibits higher PD inception voltage (PDIV) and maintains a detectable PD extinction voltage (PDEV) under reduced pressure, unlike the conventional cable. Furthermore, despite its significantly reduced thickness, the MMEI system demonstrates a substantial increase in dielectric breakdown strength, withstanding voltages exceeding 20 kV compared to below 5 kV for the conventional design under low-pressure conditions. These findings demonstrate that insulation architecture, rather than thickness alone, governs performance in MVDC aerospace cables. The results highlight the potential of MMEI systems to enable lighter, more compact, and higher-performance cable designs for future electrified aviation platforms.

## I. Introduction

THE transition toward electrified energy systems has become an important technological direction across multiple sectors, including transportation and aviation. While the pursuit of net-zero greenhouse gas (GHG) emissions remains a major motivation, it is no longer the only factor driving this shift. Similar to the transformation already seen in ground transportation, electrification in aviation is increasingly recognized as a pathway toward improved efficiency, reduced dependence on conventional fuel-based subsystems, enhanced system controllability, and the long-term modernization of aircraft power architectures. In this broader context, the transportation sector accounted for approximately 29% of total GHG emissions in the United States in 2022, while aviation contributed about 9% of that total [1]. Over the past several years, the aviation sector has continued to grow at an average annual rate of 4% to 5%, making improvements in energy utilization and aircraft system efficiency increasingly important. Narrow-body and wide-body aircraft account for more than 75% of aviation-related emissions [2-4]. At the same time, the growing interest in electrified transportation has naturally extended to aviation, where electrification is viewed not only as an environmental strategy but also as an enabling step toward next-generation aircraft technologies. Considerable progress has already been achieved in ground transportation, and similar momentum is now shaping developments in aviation.

Efforts to enhance aircraft electric power systems have led to the adoption of more-electric aircraft architectures, as exemplified by platforms such as the Boeing 787 and Airbus A380, where electrically powered subsystems increasingly replace conventional pneumatic and hydraulic systems. In addition, the emergence of hybrid-electric and all-electric aircraft concepts has further demonstrated the practical value of electrification, although current implementations remain largely limited to short-haul applications [5, 6]. The future realization of wide-body all-electric aircraft will require electric power systems with both high power density and low system mass, placing strong demands on critical components, particularly aircraft power cables.

Aircraft power cables are a critical component of electrified propulsion systems, as they are responsible for transmitting large amounts of electrical power across the airframe while meeting stringent constraints on weight, reliability, and thermal performance. For wide-body all-electric aircraft (AEA), the required electrical power is substantial. For example, studies based on the NASA N3-X aircraft indicate that power demands can exceed 25 MW during takeoff and remain on the order of

The authors are with the Department of Electrical and Computer Engineering, The University of Texas at Dallas, Richardson, TX 75080 USA (e-mail: Saikat.chowdhury@utdallas.edu; mona.ghassemi@ utdallas.edu).

several megawatts during cruise conditions [7-12]. Delivering such high power levels necessitates the adoption of medium-voltage architectures to limit conductor size and overall system mass. In this context, medium-voltage direct current (MVDC) systems in the range of several kilovolts have been identified as a practical solution, with proposed architectures commonly operating around ±5 kV [7]. At these voltage levels, cable currents can reach the order of kiloamperes, depending on system configuration and loading conditions [7, 8]. These requirements highlight the central role of power cable design in enabling high-power-density aircraft electric power systems and underscore the need for advanced insulation technologies capable of operating reliably under elevated electrical stress and harsh environmental conditions.

A natural question that arises from these requirements is whether existing MV underground or ground-based power cables can be directly adopted for wide-body AEA applications. At first glance, such an approach appears attractive, given the maturity, reliability, and extensive field experience associated with terrestrial cable technologies. However, prior studies have shown that this direct translation is not feasible due to the fundamentally different operating environment of aircraft systems. In particular, aircraft power cables must operate under significantly reduced ambient pressure (approximately 18.8 kPa at cruising altitude), where heat transfer mechanisms differ substantially from those at atmospheric conditions. As demonstrated in [13, 14], the reduced convective cooling at low pressure leads to elevated conductor and insulation temperatures, which in turn affect material conductivity and alter the electric field distribution in DC insulation systems. Beyond these thermal challenges at cruising altitude, a fundamental question arises regarding the electrical reliability of insulation systems: whether terrestrial cable designs can maintain comparable partial discharge (PD) performance and dielectric strength under reduced-pressure conditions, or whether their insulation performance degrades as ambient pressure decreases. In low-pressure environments, changes in gas density and ionization characteristics can significantly influence PD inception and propagation, potentially leading to earlier discharge activity and reduced breakdown strength. These effects raise concerns about the suitability of conventional insulation systems, originally designed for atmospheric conditions, when applied to aerospace environments. These limitations highlight the need for fundamentally different insulation design approaches tailored to MVDC aircraft applications. In particular, insulation design must move beyond thickness-driven strategies and instead target architectures that enable substantial reductions in cable mass and volume while maintaining the same voltage and current ratings—an essential requirement for aviation applications.

Multilayer insulation is a well-established concept in high-voltage engineering and has been extensively explored for a variety of applications. In aerospace systems, this concept has been further advanced with a specific focus on achieving significant reductions in insulation volume and mass while maintaining or enhancing dielectric performance. In this context, NASA research centers have developed the Micro-Multilayer Multifunctional Electrical Insulation (MMEI) system, which extends the multilayer insulation concept to meet the stringent requirements of aviation and space environments. The MMEI concept is based on stacking multiple thin dielectric layers—primarily polyimide and fluoropolymers—using controlled bonding processes to engineer the electric field distribution and enhance insulation performance. Experimental studies have demonstrated that optimized MMEI structures can achieve up to ~90% improvement in dielectric breakdown strength compared to conventional single-layer insulation, enabling nearly two orders of magnitude reduction in required insulation thickness for equivalent dielectric performance [15]. In addition, the multilayer architecture provides inherent multifunctionality, including improved corona resistance, PD mitigation, moisture resistance, and enhanced thermal durability. Full-scale validation using high-voltage busbar prototypes has further confirmed the capability of MMEI systems to operate reliably under demanding electrical stress conditions [16].

Building upon these developments, the concept of MMEI was extended to MVDC power cable design for wide-body AEA applications in [17, 18] at ±5 kV, 1 kA ratings. For DC operation, the modeling framework explicitly accounts for the strong dependence of electric-field distribution on the temperature- and field-dependent electrical conductivity of polymeric insulation materials, thereby capturing the coupled electrothermal behavior that governs DC cable performance. Within this framework, insulation is treated as a design variable rather than a fixed material choice. MMEI structures are considered using high-performance polymeric materials, including polyimide films (e.g., Kapton® MT+), fluoropolymers, and silicone-based layers. The selection of materials, layer sequencing, and thicknesses is guided by coupled electrothermal and computational fluid dynamics (CFD) simulations to satisfy the stringent electrical, thermal, and mass constraints of aircraft MVDC systems. In this approach, conductor dimensions, insulation build-up, and overall cable geometry are optimized simultaneously to achieve maximum power density with minimum mass and volume, while maintaining sufficient dielectric margins under DC stress [19, 20]. This design methodology was further extended to bipolar MVDC cable architectures, including noncircular configurations such as rectangular, square, and coaxial geometries [21–28], where conductor spacing and layout were optimized by balancing electric-field stress, thermal interaction, and geometric constraints. In addition, the surrounding installation environment was incorporated into the modeling framework, including cable ducts and confinement effects [29, 30], enabling evaluation of airflow limitations, thermal bottlenecks, and system-level heat transfer conditions that are typically neglected in conventional insulation studies. In a subsequent development, the impact of forced convection relative to natural convection on bipolar MVDC cables for wide-body AEA was systematically investigated [31, 32]. These multiphysics modeling studies of MMEI-based MVDC power cables have been further extended to other extreme conditions, such as power transmission on the lunar surface [33–36].

Importantly, these studies were not limited to modeling alone but were used to directly guide physical design and experimental validation. Optimized MMEI-based cable configurations were fabricated and tested under DC voltage stress and reduced-pressure conditions representative of aircraft environments [37–50]. Experimental results, including PD characterization and DC dielectric strength measurements, demonstrated clear performance advantages over conventional single-layer insulation systems, particularly under low-pressure operation where PD inception and breakdown mechanisms are significantly altered [47, 48]. It is important to note that, although prior comparative studies between MMEI-based and conventional underground cable systems have demonstrated the superior performance of MMEI insulation under atmospheric pressure and a low-pressure of 18.8 kPa. [48], those comparisons were not conducted under fully equivalent design conditions. In particular, differences in core conductor material (e.g., aluminum versus copper) and core conductor dimensions result in different maximum electric fields experienced by the insulation in the two cable systems. In addition, conventional underground cable features such as semiconducting conductor and insulation shields were not incorporated in the MMEI-based configurations, further limiting the direct comparability between the two systems.

In the present work, these limitations are addressed by establishing a consistent and controlled comparison framework, in which all cable components—including core conductor size, type, shielding layers, and overall configuration—are kept identical, except for the insulation system itself. Specifically, the baseline case is defined as the conventional underground cable architecture, where only its single-layer insulation is replaced by the proposed MMEI structure. This approach enables a true one-to-one comparison of insulation systems— a conventional single layer bulk insulation vs MMEI system, isolating the possible impact of other components on partial discharge behavior and dielectric strength, and thereby providing a more rigorous and physically meaningful evaluation of MMEI performance for MVDC aircraft applications.

It should be noted that the works cited, to the best of our knowledge, represent the only studies investigating MVDC power cables operating in the kV voltage and kA current range for aviation applications under low-pressure conditions (~20 kPa). These citations are not intended as self-reference, but rather reflect the fact that this body of work constitutes some of the earliest contributions in this emerging research area.

## II. Development of MMEI-Based Cable Systems

### A. Overview of MMEI Design Configurations

In earlier work, multiple MMEI system configurations were systematically developed to improve the performance of aviation power cables. The most extensively investigated designs include T-MMEI, SC-T-MMEI, ARC-SC-T-MMEI, PD-T-MMEI, and ARC-PD-T-MMEI [17].

For fabrication, three representative configurations were selected, as shown in Fig. 1. All designs share a common insulation structure consisting of three layers of Kapton® MT+ and four layers of Teflon® PFA; however, the outermost layers are modified to address specific application requirements. The PD-T-MMEI configuration is optimized for reduced mass and compactness but does not provide protection against arc tracking. To address this limitation, the ARC-PD-T-MMEI design incorporates an additional 2-mil Teflon® PFA outer layer, serving as an arc-resistant barrier. Nevertheless, both configurations lack a shielding layer, making them more susceptible to electromagnetic interference (EMI) and long-term PD effects. In contrast, the ARC-SC-T-MMEI configuration integrates a screening layer to mitigate these risks while also enhancing thermal performance through improved heat dissipation. Kapton® MT+ was selected for its high thermal conductivity and excellent dielectric performance, while Teflon® PFA was chosen for its strong fuse-bonding capability and superior resistance to arc tracking. The outermost layer, Kapton® CRC—a corona-resistant polyimide—was incorporated to enhance resistance to PD. In addition, alternative materials, including Kapton® XP and TEONEX® Q51, were evaluated during the material selection process.

Although Kapton® XP exhibits a dielectric strength of approximately 189 kV/mm at a thickness of 2 mil, and TEONEX® Q51 offers about 300 kV/mm at 0.48 mil, their unknown thermal conductivity limits their suitability for applications requiring effective thermal management. The fabrication of MMEI flat samples involves specialized equipment, including 3/16-inch A2 steel molds, C3 Conner clips for high-temperature sealing, and both forced convection and vacuum drying ovens.

The compression molds are designed with a standard length of 6 inches and are available in widths of 1, 2, and 6 inches. Figure 2 illustrates the mold design schematic, while Fig. 3 presents the fabricated ARC-SC-T-MMEI flat sample. Detailed fabrication procedures are provided in [38].

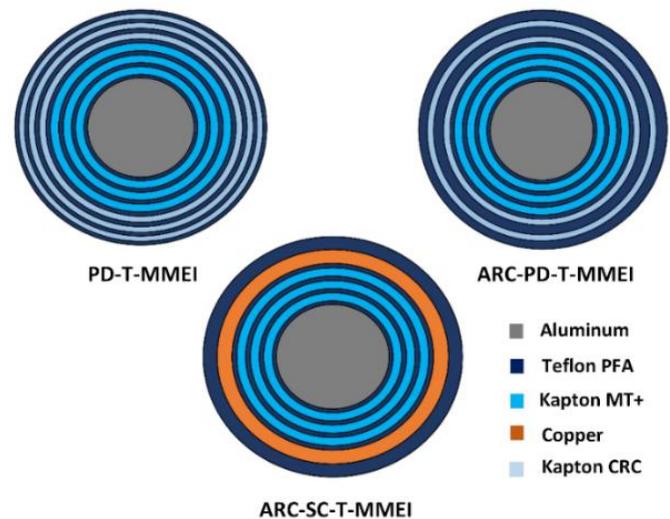

**Fig. 1.** Developed MMEI system configurations [17]. Note: The illustrations are not to scale.

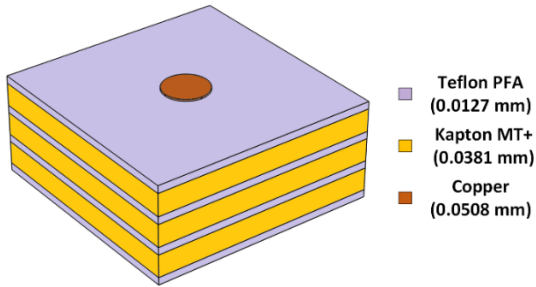

**Fig. 2.** Schematic representation of the flat MMEI sample (not to scale).

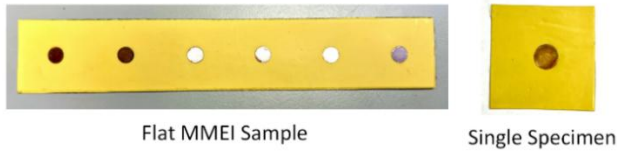

**Fig. 3.** Fabricated MMEI flat sample configuration.

To optimize the fabrication process, twelve ARC-SC-T-MMEI samples were produced under varying thermal conditions, with maximum temperatures of 320°C, 340°C, and 360°C, and holding times of 20 and 40 minutes. For half of the samples, a preliminary heating stage at 250°C for two hours in a vacuum oven was applied prior to the main thermal cycle. Analysis of PD test results indicated that the optimal fabrication condition was 360°C for 40 minutes with prior vacuum preheating [38].

Teflon PFA, selected for its high thermal stability, exhibits a melting range of approximately 302 °C to 310 °C. To ensure complete interlayer fusion while avoiding material degradation, bonding temperatures were carefully selected within a narrow window of 320 °C to 360 °C. This temperature range proved critical for achieving effective adhesion between layers. The bonding duration, varied between 20 and 40 minutes, was further optimized to balance mechanical integrity and dielectric performance.

A specially high-temperature aluminum conductor was used as the core, and a 2-mil-thick copper electrode—slightly shorter than the insulation length—was incorporated to facilitate high-voltage testing. Kapton® MT+ and Teflon® PFA films were cut into half-inch-wide strips to ensure a snug fit around the conductor. Two fabrication approaches were investigated: one employing spiraled dielectric layers applied in opposite directions, and another without spiraling. To prevent unintended bonding between the insulation and the clamping system, an additional 1.5-mil Kapton® MT+ layer was inserted between the PFA layer and the clamps. The complete assembly was then compressed and heat-treated using high-temperature clamps in a forced convection oven.

The two fabrication strategies were evaluated through PD testing, focusing on key parameters including partial discharge inception voltage (PDIV), repetitive PDIV (RPDIV), average discharge magnitude, and total PD pulse count. The results demonstrated that the spiraled configuration significantly outperformed the non-spiraled approach, achieving average PDIV and RPDIV values of 3.02 kV and 3.84 kV, respectively, compared to 2.62 kV and 3.02 kV for the non-spiraled configuration. Detailed results of this evaluation are presented in [39]. Figs. 4 and 5 illustrate the spiraled configuration, including the applied insulation layers and the final ARC-SC-T-MMEI cable fabricated using the optimized spiraling technique.

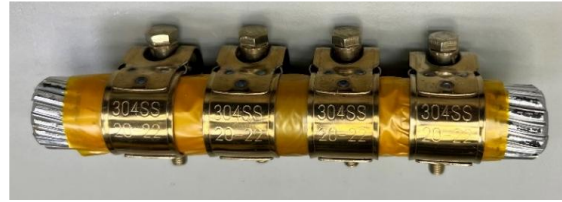

**Fig. 4.** Aluminum conductor with T-bolt hose clamps following the application of spiraled PFA and Kapton® MT+ layers.

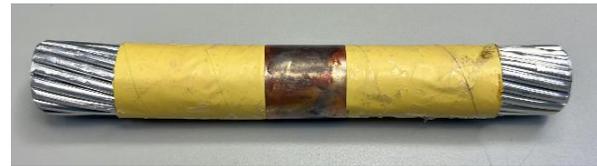

**Fig. 5.** Fabricated MMEI cable sample using the spiraled configuration.

*B. Reference Underground Cable Insulation System*

The baseline underground cable considered in this study follows a conventional shielded medium-voltage design suitable for both AC and DC applications, as illustrated in Fig. 6 [51], consisting of multiple concentric layers extending from the inner conductor to the outer jacket. The core conductor is a Class B stranded bare copper conductor, compliant with ASTM B3 and ASTM B8 standards, selected to provide an optimal balance between electrical conductivity and mechanical flexibility. No tin coating is applied in this baseline configuration. Surrounding the conductor is a semiconducting copolymer conductor shield, which serves to smooth the electric field at the conductor–insulation interface and mitigate localized field enhancement that could otherwise initiate PD.

The primary insulation layer consists of a 2.92 mm (115 mil) thick natural rubber-based ethylene propylene rubber (NL-EPR), rated for medium-voltage operation (5–8 kV). Enclosing the insulation is a strippable semiconducting copolymer insulation shield, which ensures a uniform electric field distribution, minimizes surface irregularities, and facilitates cable termination and splicing.

A helically applied copper tape shield, approximately 5 mil in thickness with about 25% overlap, is installed over the insulation shield. This metallic layer provides a low-impedance path to ground for fault and capacitive currents, enhances electromagnetic shielding, and contributes to maintaining a controlled radial electric field within the cable structure.

Finally, the outermost layer consists of a polyvinyl chloride (PVC) jacket, which provides mechanical protection against abrasion, moisture ingress, chemical exposure, and environmental degradation.

As discussed earlier, prior comparative studies [48] between MMEI-based and the above mentioned conventional underground cable system did not employ fully equivalent design conditions, limiting direct comparison. In this work, a consistent and controlled framework is established in which all cable components— including conductor type and size, shielding layers, and overall configuration—are kept identical, with the insulation system as the only variable. This enables a true one-to-one comparison between conventional single-layer insulation and the proposed MMEI architecture, allowing the impact of insulation design on partial discharge behavior and dielectric strength to be isolated and rigorously evaluated.

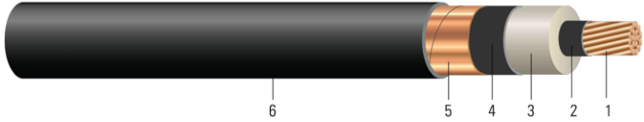

**Fig. 6.** Schematic cross-sectional representation of the baseline underground cable showing (1) stranded copper conductor, (2) semi-conducting conductor shield, (3) 115 mil NL-EPR insulation, (4) semi-conducting insulation shield, (5) copper tape shield, and (6) PVC outer jacket.

*C. Replacement of Conventional Insulation with MMEI-Based System*

To investigate the potential of insulation architecture optimization, the conventional extruded insulation layer is replaced with a MMEI system. The semiconducting conductor shield is retained to preserve proper electric field grading at the conductor–insulation interface. Following integration of the MMEI structure, the insulation shield and metallic tape shield are reapplied to maintain the original shielding and grounding functionality of the cable.

Importantly, the total insulation thickness is not preserved. Instead, the MMEI thickness is initially selected as 0.1× that of the baseline underground insulation, corresponding to 11.5 mil (0.292 mm), representing a 90% reduction in insulation thickness. The objective is to evaluate whether the engineered multilayer architecture can compensate for this substantial thickness reduction through improved electric field control and PD suppression mechanisms.

This approach enables isolation of the insulation architecture as the sole variable, while all other cable components, materials, and interfaces remain unchanged, thereby providing a rigorous and physically meaningful basis for comparison.

### III. PARTIAL DISCHARGE CHARACTERISTICS UNDER DC AND REDUCED-PRESSURE CONDITIONS

One of the primary challenges in designing aerospace insulation systems is accurately predicting their performance under the reduced pressures encountered at typical flight altitudes. Under such conditions, insulation materials can exhibit significant changes in their electrical characteristics and overall behavior. Previous studies [52, 53] have shown that as air pressure decreases below atmospheric levels, both the partial discharge inception voltage (PDIV) and breakdown voltage (BDV) are reduced. In addition, environmental factors such as temperature, pressure, and humidity strongly influence the dielectric strength of polyimide-based insulation systems. The nature of the applied voltage—whether AC or DC—further affects these properties, as partial discharge activity and breakdown mechanisms differ under alternating and direct electric fields [54].

These coupled dependencies underscore the importance of evaluating the partial discharge behavior of MMEI insulation under DC voltages across a range of pressures to assess its suitability for aerospace applications.

*A. Experimental Setup and Test Conditions*

PD responses of the MMEI-based cable samples fabricated as described in Section II.C are evaluated under both atmospheric pressure (1 atm) and reduced pressure (18.8 kPa), using the experimental setup shown in Fig. 7.

PD measurements under DC condition were conducted using a Phenix Model 4100-10 high-voltage source, capable of supplying up to 100 kV with leakage current measurement ranging from 0.01 to 20,000 µA. A 1 nF coupling capacitor rated at 150 kV was incorporated in accordance with IEC 60270 guidelines. The PD signals were measured using an OMICRON MPD 800 system, powered by an RBP1 battery pack, with current flow regulated through a high-voltage resistor.

A sphere-to-sphere electrode configuration was employed to approximate a uniform external electric field and minimize premature PD initiation due to field non-uniformities. The electrodes consisted of hemispherical terminals with a diameter of 0.5 inch (12.7 mm) and a spacing of approximately 0.25 mm. While this arrangement provides a near-uniform field, the field uniformity decreases with increasing electrode spacing.

A vacuum chamber equipped with a pumping system was used to maintain controlled pressure conditions at 1 atm and 18.8 kPa, enabling direct comparison of PD behavior under atmospheric and reduced-pressure environments. For each test, the applied voltage was increased at a rate of 1 kV/min up to 10 kV, held constant for five minutes, and then decreased at the same rate to determine the partial discharge extinction voltage (PDEV), as illustrated in Fig. 8.

The selected maximum voltage of 10 kV ensured sufficient electrical stress, particularly under low-pressure conditions. Following the identification of PD inception voltage (PDIV), the voltage was increased to 10 kV to observe sustained PD activity, typically on the order of ~20 pulses per minute. Maintaining this voltage level for five minutes enabled comprehensive analysis of both PD pulse magnitude and repetition rate.

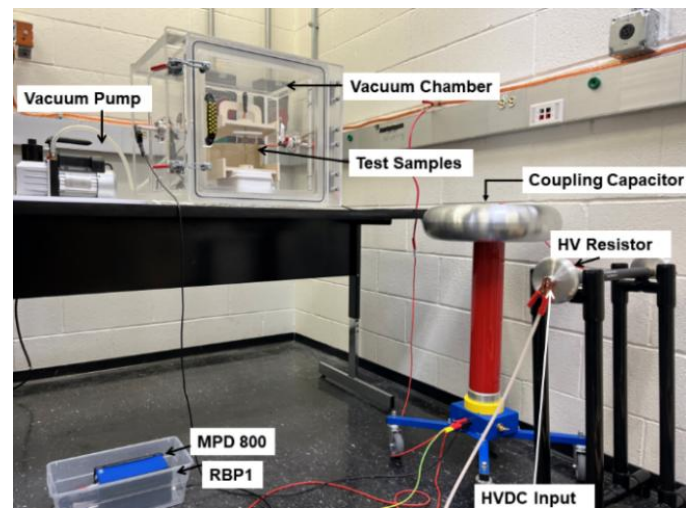

**Fig. 7.** Experimental configuration for PD measurement under DC.

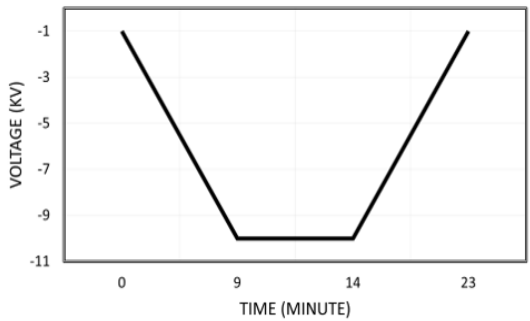
Fig. 8. Test Voltage Profile.

*B. Test Results*

Initial PD testing was conducted on the baseline underground cable samples mentioned in Section II.B, followed by evaluation of the redesigned cable samples incorporating the MMEI insulation system as mentioned in Section II.C. Tests were performed under DC voltage at atmospheric pressure and 18.8 kPa. For consistency, each cable sample was subjected to four consecutive PD measurements, and the results were averaged. Five samples were tested for the baseline underground cable, whereas two samples were fabricated and evaluated for the MMEI-based configuration.

The primary PD parameters evaluated were the PDIV and PDEV, which characterize the initiation and stability of discharge activity under both AC and DC stress conditions. Table I summarizes the averaged PDIV and PDEV values for the conventional NL-EPR cable, while Table II presents the corresponding results for the MMEI-based cable.

Because all other cable components—including conductor, shielding layers, and overall geometry—were kept identical, the comparison isolates the effect of insulation architecture on PD performance. This enables a direct assessment of the transition from conventional single-layer extruded insulation to the engineered multilayer MMEI system.

TABLE I
PARTIAL DISCHARGE CHARACTERISTICS OF THE BASELINE UNDERGROUND CABLE

| Voltage | | PDIV (kV) | PDEV (kV) |
|---|---|---|---|
| DC | 101 kPa | 1.8 | 1.3 |
| | | 1.7 | 1.4 |
| | | 1.8 | 1.2 |
| | | 1.7 | 1.2 |
| | | 1.7 | 1.4 |
| | 18.8 kPa | 1.8 | -- |
| | | 1.6 | |
| | | 1.3 | |
| | | 1.3 | |
| | | 1.3 | |

TABLE II
PARTIAL DISCHARGE CHARACTERISTICS OF THE CONVENTIONAL UNDERGROUND CABLE WITH MMEI-BASED INSULATION

| Voltage | | PDIV (kV) | PDEV (kV) |
|---|---|---|---|
| DC | 101 kPa | 2.1 | 1.5 |
| | | 2.3 | 1.3 |
| | 18.8 kPa | 1.5 | 1.1 |
| | | 1.5 | 1.1 |

*C. Analysis and Discussion*

As observed in Table I, the absence of a detectable PDEV in the conventional underground cable samples indicates that such terrestrial cable designs are not suitable for MVDC operation under low-pressure aviation conditions. This behavior can be attributed to the multilayer architecture of the MMEI system, which introduces multiple dielectric interfaces that act as barriers to charge transport and suppress the formation of continuous conductive pathways. In contrast, the homogeneous single-layer insulation in conventional cables is more susceptible to charge accumulation and localized field enhancement under low-pressure conditions, leading to earlier PD activity and the absence of stable extinction behavior.

## IV. DIELECTRIC STRENGTH

The testing procedure followed ASTM D3755 for DC breakdown evaluation [55]. Tables III and IV present the dielectric breakdown results for the conventional underground cable samples and the modified cable incorporating the MMEI insulation system, respectively.

TABLE III
DIELECTRIC BREAKDOWN STRENGTH TEST RESULTS FOR THE CONVENTIONAL UNDERGROUND CABLE

| Voltage | | Number | Breakdown Strength (kV) | Breakdown Strength (kV/mm) |
|---|---|---|---|---|
| DC | 101 kPa | 1 | 7.6 | 2.23 |
| | | 2 | 7.10 | 2.09 |
| | | 3 | 8.26 | 2.43 |
| | | 4 | 5.58 | 1.64 |
| | | 5 | 7.52 | 2.74 |
| | 18.8 kPa | 1 | 4.89 | 1.44 |
| | | 2 | 4.42 | 1.30 |
| | | 3 | 4.89 | 1.44 |
| | | 4 | 4.56 | 1.34 |
| | | 5 | 4.76 | 1.40 |

TABLE IV
DIELECTRIC BREAKDOWN STRENGTH TEST RESULTS FOR THE CONVENTIONAL UNDERGROUND CABLE WITH MMEI-BASED INSULATION

| Voltage | | Number | Breakdown Strength (kV) | Breakdown Strength (kV/mm) |
|---|---|---|---|---|
| DC | 101 kPa | 1 | 24.17 | 102.4 |
| | | 2 | 23.3 | 98.79 |
| | 18.8 kPa | 1 | 13.5 | 57.20 |
| | | 2 | 14.69 | 62.24 |

A limited number of MMEI-based cable samples (two specimens) were fabricated and tested in this study, compared to five samples for the conventional underground cable. This difference arises from the fabrication approach: while the baseline cable represents a commercially manufactured product with well-established production processes, the MMEI-based cables were manually fabricated in a laboratory environment using a multi-step layering and bonding procedure. This process is inherently both time-intensive and challenging, and does not yet benefit from automated or industrial-scale manufacturing.

Despite the limited sample size, the experimental results demonstrate a clear and consistent performance advantage of the MMEI-based insulation system. In terms of partial discharge behavior, the MMEI configuration exhibits higher PDIV and, importantly, maintains a detectable PDEV even under reduced pressure (18.8 kPa), in contrast to the conventional cable where no PDEV is observed. This indicates improved discharge stability and resistance to the formation of conductive pathways under low-pressure conditions.

More significantly, dielectric breakdown results reveal a substantial enhancement in insulation performance. While the conventional underground cable exhibits breakdown at voltages below 5 kV under reduced pressure, the MMEI-based cable withstands voltages exceeding 20 kV, despite having only 10% of the insulation thickness. This corresponds to an order-of-magnitude increase in breakdown strength (kV/mm), clearly demonstrating the effectiveness of the engineered multilayer architecture.

Although a limited number of samples were tested, the magnitude of improvement (>3× in breakdown voltage and >10× in breakdown strength) significantly exceeds expected experimental variability, supporting the robustness of the observed trends.

It is important to emphasize that the presented MMEI samples represent early-stage, manually fabricated prototypes. With the adoption of controlled, automated manufacturing processes, further improvements in interfacial quality, defect minimization, and overall dielectric performance are expected. Therefore, the observed performance gains are not only significant but also conservative, reinforcing the strong potential of MMEI insulation for MVDC aerospace cable applications.

## V. Conclusion

This work presented a controlled experimental comparison between conventional underground cable insulation and a micro-multilayer multifunctional electrical insulation (MMEI) system for MVDC cable applications under low-pressure conditions relevant to all-electric aircraft. By maintaining identical cable configurations and varying only the insulation architecture, the study isolated the impact of insulation design on partial discharge behavior and dielectric strength. Experimental results demonstrate that conventional single-layer insulation exhibits degraded performance under reduced pressure, including the absence of detectable partial discharge extinction voltage (PDEV), indicating early formation of conductive pathways and reduced reliability. In contrast, the MMEI-based insulation system shows improved PD performance, including higher PD inception voltage (PDIV) and sustained PDEV under low-pressure conditions. More significantly, the MMEI system achieves a substantial enhancement in dielectric strength despite a 90% reduction in insulation thickness. Under reduced pressure (18.8 kPa), the MMEI-based cable withstands voltages exceeding 20 kV, whereas the conventional cable fails below 5 kV. This corresponds to an order-of-magnitude improvement in breakdown strength (kV/mm), highlighting the effectiveness of engineered multilayer insulation in controlling electric field distribution and suppressing discharge activity. Although a limited number of MMEI samples were tested due to the complexity of manual fabrication, the magnitude and consistency of the observed improvements clearly demonstrate the superiority of the multilayer architecture. Nevertheless, the magnitude of improvement—exceeding a factor of three in breakdown voltage and an order-of-magnitude increase in breakdown strength—significantly exceeds expected experimental variability, supporting the robustness of the observed trends. Furthermore, these results represent a conservative estimate of performance, as future implementation using automated manufacturing processes is expected to further enhance insulation quality and reliability. Overall, this work establishes that insulation architecture, rather than thickness alone, is a critical design parameter for MVDC cable systems in aerospace applications. The proposed MMEI approach offers a promising pathway toward lightweight, high-performance, and reliable cable systems for next-generation electrified aircraft.

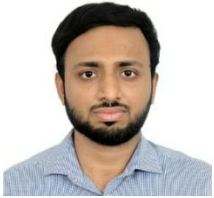

**Saikat Chowdhury** (Member, IEEE) received his B.S. degree from Bangladesh University of Engineering and Technology (BUET), Bangladesh. He is pursuing a Ph.D. in electrical engineering at The University of Texas at Dallas, TX, USA. His primary research focuses on electrical insulation for transportation electrification.

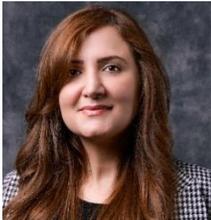

**Mona Ghassemi** (Senior Member, IEEE) received the Ph.D. degree (Hons.) in electrical engineering from the University of Tehran, Tehran, Iran, in 2012. From 2013 to 2015, she was a Postdoctoral Fellow with the NSERC/ Hydro-Québec/UQAC Industrial Chair on Atmospheric Icing of Power Network Equipment (CIGELE), University of Québec at Chicoutimi (UQAC), Canada. She has been a Registered Professional Engineer since 2015. From 2015 to 2017, she held a postdoctoral position with the University of Connecticut, Storrs, CT, USA. In 2017, she joined the Bradley Department of Electrical and Computer Engineering, Virginia Tech, as an Assistant Professor. In 2021, she was named both the Steven O. Lane Junior Faculty Fellow and the College of Engineering Faculty Fellow at Virginia Tech. In 2022, she joined the Department of Electrical and Computer Engineering, The University of Texas at Dallas (UT Dallas), as an Associate Professor, and was appointed Chairholder of the Texas Instruments Early Career Award. Since 2025, she has been a Professor at UT Dallas. She has authored more than 250 peer-reviewed journal and conference papers, and four book chapters. Her research interests include dielectrics and electrical insulation, high voltage engineering, power systems, and plasma science. She serves as the Vice President −Administrative of the IEEE Dielectrics and Electrical Insulation Society (DEIS); a DEIS Representative on the IEEE-USA Technology Policy Council's Research and Development Policy Committee; an Active Member of several CIGRÉ Working Groups and IEEE Task Forces; and a Technical Committee Member on Dielectrics and Electrical Insulation for Transportation Electrification. She also serves on the Education Committees of both IEEE DEIS and the IEEE Power and Energy Society (PES). Her previous service includes roles as Vice President−Technical of IEEE DEIS, a member of the Nominations and Appointments Committee and an At-Large Member of the Administrative Committee of IEEE DEIS, and a DEIS Representative to the IEEE-USA Committee on Transportation and Aerospace Policy (CTAP). She was a recipient of three of the most competitive, prestigious early-career awards in the United States: U.S. Department of Energy (DOE) Early Career Research Program Award, the National Science Foundation (NSF) CAREER Award, and the Air Force Office of Scientific Research (AFOSR) Young Investigator Research Program (YIP) Award. She has also received a Contribution Award from IET High Voltage and four Best Paper Awards. She serves as an Associate Editor for IEEE Transactions on Dielectrics and Electrical Insulation, IEEE Transactions on Industry Applications, IET High Voltage, International Journal of Electrical Engineering Education, and Power Electronic Devices and Components. She is also a Guest Editor for Aerospace and Energies, and an Associate Guest Editor for the IEEE Journal of Emerging and Selected Topics in Power Electronics and IEEE Transactions on Power Electronics.